\documentstyle[epsf]{ptptex}

\newcommand {\ee}{\end{equation}}
\newcommand {\bea}{\begin{eqnarray}}
\newcommand {\eea}{\end{eqnarray}}
\newcommand {\nn}{\nonumber \\}
\newcommand {\Tr}{{\rm Tr\,}}

\newcommand {\m}{\mu}
\newcommand {\n}{\nu}
\newcommand {\pl}{\partial}
\newcommand {\p} {\phi}

\newcommand {\al}{\alpha}
\newcommand {\be}{\beta}
\newcommand {\ga}{\gamma}
\newcommand {\Ga}{\Gamma}

\newcommand {\th}{\theta}

\newcommand {\om}{\omega}
\newcommand {\Om}{\Omega}
\newcommand {\ep}{\epsilon}

\newcommand {\del}  {\delta}
\newcommand {\Del}  {\Delta}
\newcommand {\mn}{{\mu\nu}}

\newcommand {\half}{ {\frac{1}{2}} }
\newcommand {\fourth} {\frac{1}{4} }

\newcommand {\Lcal}{{\cal L}}

\newcommand {\Dcal}{{\cal D}}

\newcommand {\Dvec}{{\hat D}}   

\newcommand {\Dhat}{{\hat D}}

\newcommand {\Sbar}{{\bar S}}
\newcommand {\psibar}{{\bar \psi}}
\newcommand {\chibar}{{\bar \chi}}
\newcommand {\ra} {\rightarrow}

\newcommand {\pr}   {{\quad .}}
\newcommand {\com}  {{\quad ,}}
\newcommand {\q}    {\quad}

\def\overleftrightarrow#1{\vbox{\ialign{##\crcr
 $\leftrightarrow$\crcr\noalign{\kern-1pt\nointerlineskip}
 $\hfil\displaystyle{#1}\hfil$\crcr}}}

\def\leftdslash{{\overleftarrow\dslash}}

\newcommand {\gago}  {\gamma_5}
\newcommand {\Ktil}  {{\tilde K}}

\newcommand {\Kbar}  {{\bar K}}
\newcommand {\GfMp}  {G^{5M}_+}

\newcommand {\GfMm}  {G^{5M}_-}
\def\Aslash{{}\hbox{\hskip2pt\vtop
 {\baselineskip23pt\hbox{}\vskip-24pt\hbox{/}}
 \hskip-11.5pt $A$}}
\def\aslash{{}\hbox{\hskip2pt\vtop
 {\baselineskip23pt\hbox{}\vskip-24pt\hbox{/}}
 \hskip-10.5pt $a$}}
\def\kslash{{}\hbox{\hskip2pt\vtop
 {\baselineskip23pt\hbox{}\vskip-24pt\hbox{/}}
 \hskip-8.5pt $k$}}
\def\dbslash{{}\hbox{\hskip2pt\vtop
 {\baselineskip23pt\hbox{}\vskip-24pt\hbox{$\backslash$}}
 \hskip-11.5pt $\partial$}}
\newcommand {\dslash} { {\not\partial}}  

\def\Ktilbslash{{}\hbox{\hskip2pt\vtop
 {\baselineskip23pt\hbox{}\vskip-24pt\hbox{$\backslash$}}
 \hskip-11.5pt ${\tilde K}$}}

\def\Kbarbslash{{}\hbox{\hskip2pt\vtop
 {\baselineskip23pt\hbox{}\vskip-24pt\hbox{$\backslash$}}
 \hskip-11.5pt ${\bar K}$}}

\nofigureboxrule
\preprintnumber[3cm]{
Oct. 2001\\ RIMS-1336}

\title{
Renormalization\\
 using \\
Domain Wall Regularization
}

\author{
Shoichi {\sc Ichinose}\footnote{
Permanent address:\   
Laboratory of Physics, School of Food and Nutritional Sciences, University of Shizuoka,
Yada 52-1, Shizuoka 422-8526, Japan.\\
E-mail:\ ichinose@u-shizuoka-ken.ac.jp} 
}

\inst{
Research Institute for Mathematical Sciences\\
Kyoto University, Kyoto 606-8502, Japan
}

\recdate{
}

\abst{
We formulate the renormalization
procedure using the domain wall regularization
that is based on the heat-kernel method. 
The quantum effects of both fermions and bosons (gauge fields)
are taken into account. 
The background field
method is quite naturally introduced.
With regard to the treatment of the 
loop-momentum integrals,
an interesting contrast between the fermion-determinant part
and other parts is revealed.
These points are elucidated by considering some examples.
The Weyl anomalies for 2D QED and 4D QED are correctly
obtained. It is found that
the ``chiral solution'' produces (1/2)$^{d/2}$ $\times$
``correct values'', where $d$ is the spatial dimension.
Considering the model of 2D QED, both
Weyl and chiral anomalies are directly obtained from the effective action.
The mass and wave function renormalization
are explicitly performed in 4D QED. 
We confirm the multiplicative (not additive)
renormalization, which demonstrates the
advantage of no fine-tuning. The relation with
the recently popular 
higher-dimensional approach, such as the Randall-
Sundrum model, is commented on.
}

\begin{document}

\maketitle
\section{Introduction}
In the lattice world, it appears as if a new era is about to begin.
\cite{Neu99Talk,Creu99}
The so-called ``doubling problem'' has become clarified, and
the chirality of fermions can be controlled more efficiently than before.
We may say a new regularization has been established in the lattice.
In accordance with these great developments 
, the counterpart in the continuum
is progressing.
\cite{NN94,NN95,DS95NP,DS95PL,DS97PL}  There are two 
merits to such an approach. First, we can 
clearly understand
the essential part of the new regularization, 
which is often hidden in the complexity of the discretized model. 
Second, we can compare the new regularization
with the ordinary ones used to this time
and apply it to various (continuum) field theories.
The continuum version, at least,  should explain the qualitative features
of the lattice domain wall.

In the domain wall approach used to now, at least in the continuum approach,
the gauge field in the fermion determinant is treated mainly as 
an external field.
(For lattice model analysis, the gauge field quantum effect
was examined, and the renormalizability was checked at the 1-loop level in
Ref.\cite{Yama98}.) 
Clearly, the situation is not satisfactory, because the gauge field
is not treated as a quantum field and 
the general perspective regarding its use in (perturbative) quantum field theory 
has not been
presented. Of course, the fermion determinant could be
the most important among the other parts, but we must formulate
the gauge field within the general setting in order to regard it as a new regularization
in the field theory. This is what motivates the present work.

Recently, a new treatment has appeared 
in the continuum approach.\cite{SI98,SI99} 
The main idea is the following. 
The (regularization) ambiguity and the divergences of the
fermion determinant can be resolved by introducing a ``direction"
into the system. This is based on an analogy to the well-defined
determinant of the elliptic operator, which can be expressed
in the form of a heat equation solution. 
%
%
Here we recall two facts:\ 
1)
the heat propagates in a fixed direction, that is, from
high temperature to low temperature (as stipulated by 
the second law of
the thermodynamics) and, generally, the {\it heat equation}
describes such behaviour;
2)
the famous procedure of introducing
the heat equation into a general quantum system is the heat-kernel
method.\cite{Sch51} 
The fermion determinant is very often examined using this method.
(The anomaly is formulated from this point of view in Ref.\cite{II96}.)
We have shown that the above idea works well if we consider the 4 
dimensional theory
from the 1+4 dimensional space-time.
 
We summarize this new treatment in \S 2. 
New results regarding 2D and 4D QED Weyl anomalies are also presented.
In \S 3, the renormalization
procedure is introduced, and the (renormalized) effective action 
itself is used to derive anomalies. At this stage, we still keep
the gauge field external, in order to make the explanation simple.
In \S 4, we introduce the background
field method\cite{tH73,IO82} and 
take into account the quantum effect of the gauge field
as well as that of the fermion.
We do the renormalization of the fermion self-energy explicitly.
In \S 5, we present the complete formulation, combining
the results of \S\S 3 and 4. Finally, we point out that
the domain wall regularization
treats the fermion determinant part distinctly 
from the other parts.

\section{Domain wall regularization}
For a fermion system described by the quadratic form
$\Lcal=\psibar\Dvec\psi$, 
where the operator $\Dhat$ satisfies
\begin{eqnarray}
\ga_5\Dvec+\Dvec\ga_5=0 \com\q (\gago)^2=1\com
\label{DWR1}
\end{eqnarray}
the fermion determinant (the effective action)
can be expressed as
\begin{eqnarray}
\ln\,Z[A_\m]=-\int_0^{\infty}\frac{dt}{t}\Tr\left[
\half(1+i\ga_5)\exp\{it\ga_5\Dvec\}+
\half(1-i\ga_5)\exp\{-it\ga_5\Dvec\}\right]\nn
=-\lim_{M\ra 0}\int_0^{\infty}\frac{dt}{t}
\half(1-i\frac{\pl}{\pl(tM)})\Tr ( \GfMp(x,y;t)+
                                   \GfMm(x,y;t) )
  \ ,
\label{DWR2}
\end{eqnarray}
where $t$ plays the role of the inverse temperature. Here  
$M$ is introduced as a {\it regularization} mass parameter
{\bf [Regularization Prescription 1]}.
\footnote{
Along the flow of the regularization description, the steps 
are numbered as Regularization Prescription 1,2,3,3' and 4.
} 
$M$ also plays the role of the source for $\ga_5$. 
$A_\m$ represents the background gauge field appearing in
$\Dhat$. 
We take $M>0$ for simplicity.
$\GfMp$ and $\GfMm$ are defined as
\begin{eqnarray}
\GfMp(x,y;t)= <x|\exp\{+it\ga_5(\Dvec+iM)\}|y>\com\nn
\GfMm(x,y;t)= <x|\exp\{-it\ga_5(\Dvec+iM)\}|y>
\com
\label{DWR3}
\end{eqnarray}
and they satisfy the {\it heat equations}
with the {\it first} derivative operator
\ $\mp i\gago (\Dhat+iM)$.
We call $\GfMp$ and $\GfMm$ the ``(+)-domain'' and ``(-)-domain'', 
because in the extreme chiral limit (defined below) they have
exponential distributions in the extra $t$-space, peaked
at $t=0$ or $t=\infty$. 
\footnote{
See Figs.2-4 of Ref.\cite{SI99}.
The symmetric solutions (defined below) have double peaks,
whereas the ``chiral solutions" (defined below) have a single peak.
}. 
The key observation is that the heat equations turn out to be 
the 1+4 dimensional 
Minkowski {\it Dirac equation} after appropriate {\it Wick rotations} for $t$.
For the system of 4 dimensional Euclidean QED,
$\Dvec=i\ga_\m(\pl_\m+i e A_\m),\ (\m=1,2,3,4)$, they are
\begin{eqnarray}
(i\dbslash-M)\GfMp=i e \Aslash \GfMp\com\q
(X^a)=(-it,x^\m)\com                    \nn
(i\dbslash-M)\GfMm =i e \Aslash \GfMm\com\q
(X^a)=(+it,x^\m)
\com
\label{DWR4}
\end{eqnarray}
where $\Aslash\equiv\ga_\m A_\m(x)\ ,\ 
\dbslash\equiv\Ga^a\frac{\pl}{\pl X^a}\ (a=0,1,2,3,4)$.
\footnote{
The slash ``$/$" is used for 4 dimensional contraction, whereas
the ``backslash" $\backslash$ is used for 5 dimensional contraction. 
}
\footnote{
The massive case, $\Dvec_m=\Dvec-m$ (where $m$ is the 4D fermion mass), 
is treated similarly. The equations in (\ref{DWR4})
are replaced by
$(i\dbslash-M){G^{5M}_\pm}=(i e \Aslash\mp\gago m){G^{5M}_\pm}$.
(See Ref.\cite{SI99}.)
}
Now, we can {\it specify} the above solution based on another
key observation that the system should have a {\it fixed direction}
{\bf [Regularization Prescription 2]}.
Generally, the solution of (\ref{DWR4}) is given by two ingredients
(see a standard textbook, for example, Ref.\cite{BD}), 
1) a free solution $G_0(X,Y)$,\ and 2) a propagator $S(X,Y)$\ 
in the following form:
\begin{eqnarray}
G^5_M(X,Y)=G_0(X,Y)+\int d^5Z\,S(X,Z)i e\Aslash(z)G^5_M(Z,Y)
\pr
\label{DWR4B}
\end{eqnarray}
This gives us the solution $G^5_M(X,Y)$ perturbatively
with respect to the coupling $e$.
As $S(X,Y)$, we {\it should not} use the Feynman propagator 
that has both retarded
and advanced components. 
Instead we should use the retarded propagator
for the (+)-domain and the advanced propagator for the (-)-domain:
\begin{eqnarray}
\mbox{Symmetric retarded solution}\q \GfMp
 (\mbox{(+)-domain}):\q \nn
G_0(X,Y)=G^{p}_0(X,Y)-G^{n}_0(X,Y)\ ,\nn 
S(X,Y)=\th (X^0-Y^0)(G^{p}_0(X,Y)-G^{n}_0(X,Y))\label{DWR5A}\\
\mbox{Symmetric advanced solution}\q \GfMm
 (\mbox{(-)-domain}):\q \nn
G_0(X,Y)=-G^{p}_0(X,Y)+G^{n}_0(X,Y)\ ,\nn 
S(X,Y)=\th (Y^0-X^0)(-G^{p}_0(X,Y)+G^{n}_0(X,Y))\ ,
\label{DWR5B}
\end{eqnarray}
where $G^{p}_0(X,Y)$ and $G^{n}_0(X,Y)$ are 
the positive and negative energy free solutions, respectively:
\begin{eqnarray}
G^p_0(X,Y)\equiv -i\int\frac{d^4k}{(2\pi)^4}\Om_+(k) e^{-i\Ktil(X-Y)}\com\ \ 
\Om_+(k)\equiv\frac{M+\Ktilbslash}{2E(k)}\com\nn
G^n_0(X,Y)\equiv -i\int\frac{d^4k}{(2\pi)^4}\Om_-(k) e^{+i\Kbar(X-Y)}\com\ \ 
\Om_-(k)\equiv\frac{M-\Kbarbslash}{2E(k)}\com
\label{DWR6}
\end{eqnarray}
where $E(k)=\sqrt{k^2+M^2},(\Ktil^a)=(E(k),K^\m=-k^\m),
(\Kbar^a)=(E(k),-K^\m=k^\m)$. 
Here, $k^\m$ is the momentum in the 4 dimensional Euclidean space,
\footnote{
The relations between 4 dimensional quantities\ ($x^\m$ and $k^\m$)
and 1+4 dimensional quantities\ ($X^a$ and $K^a$) are as follows :\ 
$(X^a)=(X^0,X^\m=x^\m), (K^a)=(K^0,K^\m=-k^\m),
K_aX^a=K_0X^0-K^\m X^\m=K_0X^0+k^\m x^\m.$
}
and $\Ktil$ and $\Kbar$ are the on-shell
momenta, $\Ktil^2=\Kbar^2=M^2$, which correspond to the positive
and negative energy states, respectively.
\footnote
{The following are 
useful relations:\ 
$
-i\Ktil X=-iE(k)X^0-ikx,\ i\Kbar X=iE(k)X^0-ikx,\ 
M+\Ktilbslash=M+E(k)\gago+i\kslash,\ 
M-\Kbarbslash=M-E(k)\gago+i\kslash.
$
}
The theta functions in (\ref{DWR5A}) and (\ref{DWR5B}) show
the ``directedness'' of the solution.
In this solution, both the positive and negative states
propagate in the forward direction in 
the (+)-domain, while they both propagate
in the backward direction in the (-)-domain.
We call (\ref{DWR5A}) and (\ref{DWR5B}) ``symmetric solutions''. 
(They are ``symmetric" in the sense that the positive and negative
energy parts are equally mixed. Compare them with
the chiral ``solution" presented below. )
The above solutions satisfy the following 
{\it boundary conditions}:
\begin{eqnarray}
\GfMp\mbox{(Retarded)}\ra -i\gago\del^4(x-y)\ \mbox{as}\ 
M(X^0-Y^0)\ra +0\ ,\nn
\GfMm\mbox{(Advanced)}\ra +i\gago\del^4(x-y)\ \mbox{as}\ 
M(X^0-Y^0)\ra -0\ .
\label{DWR7}
\end{eqnarray}

In this procedure, we naturally notice 
the following attractive choice of $G_0(X,Y)$ and $S(X,Y)$:
\footnote{
In Refs.\cite{SI98} and \cite{SI99}, we called 
the ``solution" obtained by this choice the 
``Feynman path solution'', 
because it was ``invented'' by ``dividing'' the Feynman propagator.
}
\begin{eqnarray}
\mbox{Retarded chiral "solution"}\q \GfMp
 (\mbox{(+)-domain}) :\q \nn
G_0(X,Y)=G^p_0(X,Y)\com\q S(X,Y)=\th (X^0-Y^0)G^p_0(X,Y)\label{DWR8A}\\
\mbox{Advanced chiral "solution"}\q \GfMm
 (\mbox{(-)-domain}) :\q \nn
G_0(X,Y)=G^n_0(X,Y)\com\q S(X,Y)=\th (Y^0-X^0)G^n_0(X,Y)\pr
\label{DWR8B}
\end{eqnarray}
They also represent ``directed'' propagations, but
do {\it not} satisfy (\ref{DWR4}). Instead they satisfy
its {\it chiral version} in the large $M$ limit 
or the soft-momentum limit ($M/|k^\m|\gg 1$),
\begin{eqnarray}
(i\dbslash-M)\GfMp=i e P_+\Aslash \GfMp+O(\frac{1}{M})\com\q
(X^a)=(-it,x^\m)\com                    \nn
(i\dbslash-M)\GfMm =i e P_-\Aslash \GfMm+O(\frac{1}{M})\com\q
(X^a)=(+it,x^\m)
\pr
\label{DWR9}
\end{eqnarray}
The configuration in which the positive energy states propagate
only in the forward direction of $X^0$ constitutes the ($+$)-domain,
while the configuration in which the negative energy states propagate 
only in the backward direction constitutes the ($-$)-domain.
As seen from their simple structure, 
the chiral ``solutions" have some advantages (at least in concrete calculations).
The validity of their use, however, is a subtle matter, because
they are legitimate solutions of {\it neither} (\ref{DWR4}) 
{\it nor} its chiral version. Only in the {\it extreme chiral}
limit ($M/|k|\gg 1$) do they correspond to the chiral operators
$\Dvec_\pm =i(\dslash+i e P_\pm\Aslash)$. 
We believe that their use is valid in the examination of
global quantities or soft-momentum (infrared) properties.
(This will be confirmed below.) 
The chiral ``solutions" satisfy the {\it boundary condition}
\begin{eqnarray}
i(\GfMp(X,Y)-\GfMm(X,Y))\ra \gago\del^4(x-y)\ \mbox{as}\ 
M|X^0-Y^0|\ra +0\pr
\label{DWR10}
\end{eqnarray}

As the final step of this regularization prescription, we take 
the following ``double limits''{\bf [Regularization Prescription 3]}:
\begin{eqnarray}
\mbox{(i)}\q \frac{|k^\m|}{M}\leq 1\com\q
\mbox{(ii)}\q Mt\ll 1\com\nn
(\mbox{or}\q |k^\m|\leq M\ll \frac{1}{t}) \pr
\label{DWR11}
\end{eqnarray}
These relations express the most characteristic point of this
1+4 dimensional regularization scheme.
The condition (ii) comes from the usage of  
the regularization parameter $M$
in (\ref{DWR2}), whereas (i) comes from the necessity of controlling
the chirality without destroying the system dynamics.
\footnote{
In the {\it extreme chiral} limit $M/|k|\gg 1$, $\Om_\pm(k)$
 converge to the chiral projection operators: 
$\Om_\pm(k)\ra P_\pm$ .
The control of the chirality is 
regarded as complete in this limit.
The restriction, however, is too strong to maintain the dynamics.
Hence, we consider the ``loosened" restriction (i). 
This procedure
should be regarded as a part of the present regularization.
}
Note that the roles of the regularization mass parameter $M$
for the $t$-axis and for the $x^\m$-axis are different.
The parameter $M$ restricts the configuration
to the {\it ultra-violet} region ($t\ll M^{-1}$)
in the ``extra space'' of $t$ and
to the {\it infra-red} (surface) region in the 
real 4 dimensional space ($|k^\m|\leq |M|$). 
\footnote{
This point is what we mention, at item 6) [infrared-ultraviolet
relation] of the final paragraph of \S 5, as an analogous phenomenon
to the brane-world approach.
}
(This situation regarding configuration restriction, by (i) and (ii), 
is discussed further in \S\S 3 and 5.) 
In the concrete calculation,
the condition (i) is taken into account 
not by performing $k^\m$-integral with the cut-off $M$
but by using the {\it analytic continuation} in order to avoid breaking the gauge invariance{\bf [Regularization Prescription 3']}.
\footnote{
The explicit use of the momentum cut-off $M$ ($|k^\m|\leq M$)
clearly breaks gauge invariance. We can show that an equivalent
regularization is realized by analytic regularization, 
where the cut-off parameter is not necessary. (See Ref.\cite{SI99}.)
}

The validity of the above regularization was confirmed in 
Refs.\cite{SI98} and \cite{SI99}. There
we found properties analogous to those of the lattice domain wall:\ 
the domain wall configuration, the overlap
Dirac operator, the condition on the regularization parameter $M$, etc..
We also confirmed that the chiral anomalies (for 2D QED, 4D QED and
2D chiral gauge theory) are correctly reproduced.
One of the advantages of the present approach is the equal treatment
of the {\it chiral} and {\it Weyl} anomalies.
To understand the situation, 
let us apply the above regularization to the Weyl anomaly
calculation.
We first consider a simple model, 2D QED, for later purposes. 
It is given, using the chiral solution (\ref{DWR10}),  as\cite{SI99}
\begin{eqnarray}
\del_\om\,\ln\,J_W=2i\lim_{M|X^0-Y^0|\ra +0}
\Tr\,\om(x)\gago (\GfMp(X,Y)-\GfMm(X,Y))\com\nn
Tr\,\om\gago\GfMp|_{AA}
=\int^{X^0}_0dZ^0\int d^2Z\int^{Z^0}_0dW^0\int d^2W\nn
\times \Tr\,\om\gago G^p_0(X,Z)i e \Aslash(z)
G^p_0(Z,W)i e \Aslash(w)G^p_0(W,Y)
\com
\label{DWR12}
\end{eqnarray}
(See Fig.1(i).)  and 
similarly for $\mbox{Tr}\,\om\gago\GfMm$ using $G^n_0$. After
a standard calculation explained in Ref.\cite{SI99}, we obtain
{\it half} of the correct result:
$\del_\om\,\ln\,J_W=\om(x)\half\frac{e^2}{\pi}{A_\m}^2$. When we take
the symmetric solution, we evaluate
\begin{eqnarray}
\del_\om\,\ln\,J_W=i\lim_{M|X^0-Y^0|\ra +0}
\Tr\,\om(x)\gago (\GfMp(X,Y)-\GfMm(X,Y))\com\nn
Tr\,\om\gago\GfMp|_{AA}
=\int^{X^0}_0dZ^0\int d^2Z\int^{Z^0}_0dW^0\int d^2W\nn
\times\Tr\,\om\gago (G^p_0(X,Z)-G^n_0(X,Z))i e \Aslash(z)
(G^p_0(Z,W)-G^n_0(Z,W))\nn
\times i e \Aslash(w)(G^p_0(W,Y)-G^n_0(W,Y))
\pr
\label{DWR13}
\end{eqnarray}
(See Fig.1(ii).) This reproduces  the correct result:
$\del_\om\,\ln\,J_W=\om(x)\frac{e^2}{\pi}{A_\m}^2$.
%
\begin{figure}
\centerline{\epsfysize=6cm\epsfbox{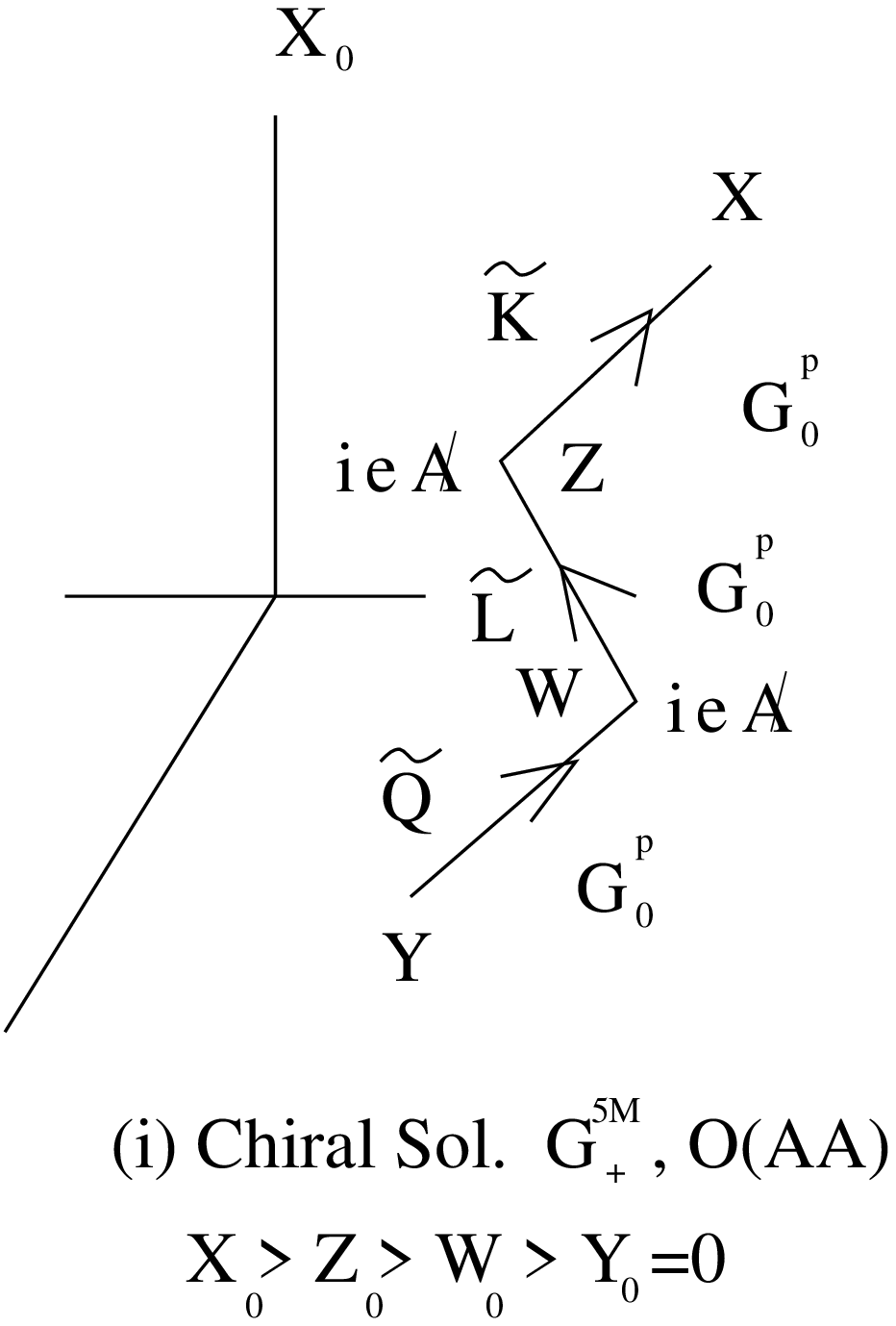}\\
\epsfysize=6cm\epsfbox{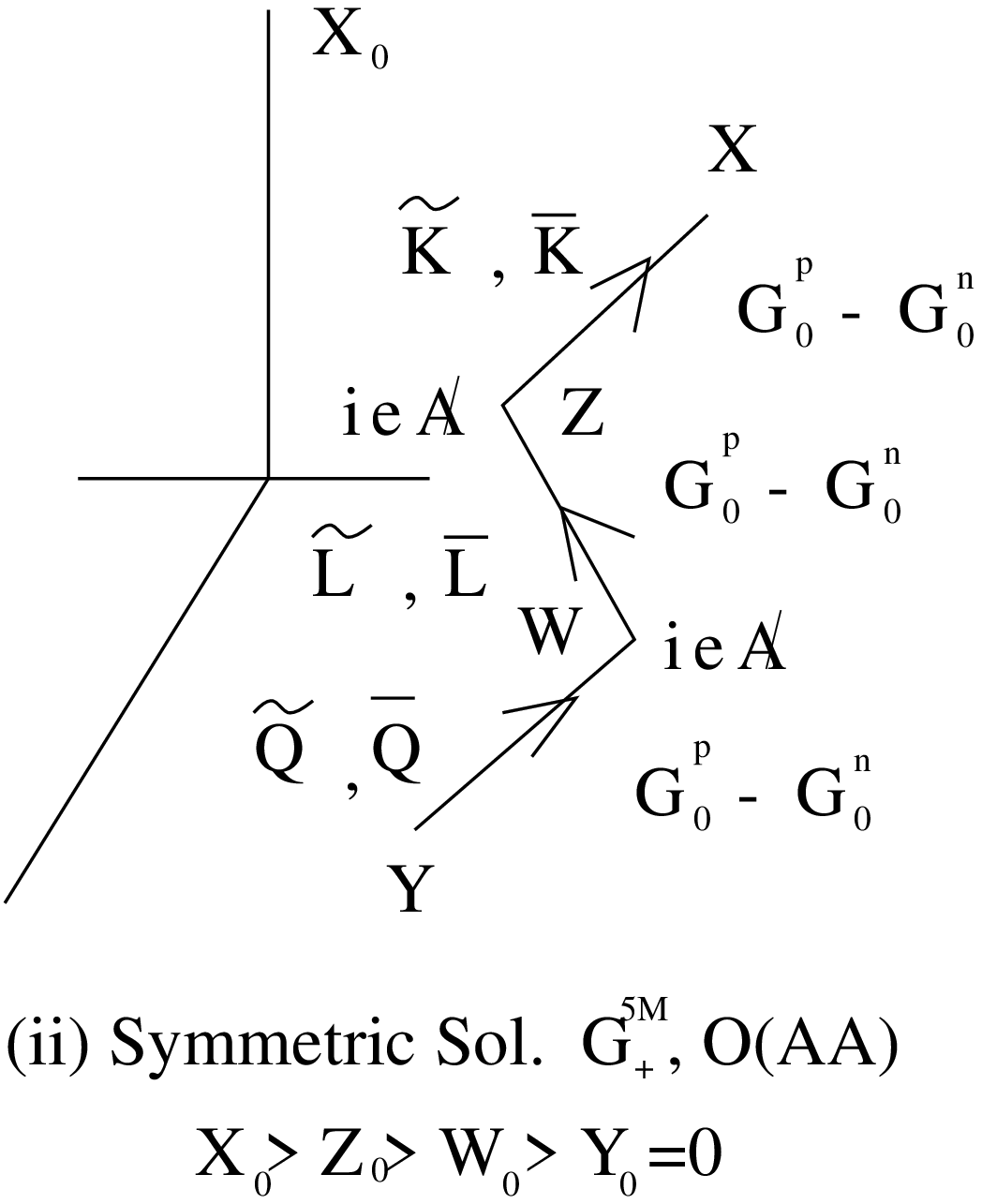}}
\caption{
Abelian gauge theory, $\GfMp$, 
second order with respect to $A_\m$
[$O(AA)$]: (i) the chiral ``solution" and
(ii) symmetric solution.}
\end{figure}
%
We have confirmed that similar situation exists
in 4D Euclidean QED. The symmetric
solution gives the correct Weyl anomaly:\ 
$\del_\om\ln J=\om(x)\,\be(e)F_\mn F_\mn \ ,\ 
\be(e)=\frac{e^2}{12\pi^2}$, where $\be(e)$ is the $\be$-function
in the renormalization group.
\footnote{
The general anomaly calculation based on
the ordinary (i.e. that in which the domain wall is not used) 
heat-kernel is reviewed in Ref.\cite{SI99}.
}
The chiral ``solution" gives 
{\it one fourth} of it.   

Combined with the results for the chiral anomalies in 
Refs.\cite{SI98} and \cite{SI99},
we conclude that 
the chiral ``solution" gives $(1/2)^{d/2}$ (where $d$ is the 
spatial dimension) 
of the correct value of the anomaly coefficients
both for the chiral anomaly and for the Weyl anomaly.
The use of the chiral ``solution" reduces the degree of freedom
by half for each two dimensions.
This phenomenon seems to contrast with the lattice's
doubling species phenomenon.

The use of the chiral ``solution", instead of the symmetric solution, 
should be limited to 
simple cases, such as the anomaly calculation. 
For the general case, in which local dynamics appears
and we cannot ignore $O(1/M)$ in (\ref{DWR9}), 
we should use the symmetric solution.

\section{Renormalization procedure}
In this and the next sections, we develop a method for using
the domain wall regularization in general field theory. 
We now introduce the {\it counterterm} action $\Del\Ga[A]$ into the
effective action $\Ga[A]\equiv -\ln Z[A]$ as
\begin{eqnarray}
\Ga_{\mbox{bare}}[A]\equiv \Ga[A]+\Del\Ga[A]\com
\label{REN1}
\end{eqnarray}
in such a way that $\Ga_{\mbox{bare}}$ becomes finite. Determining 
how to systematically define $\Del\Ga$ and obtain
the renormalization properties is the task of this section.
Let us consider the simple model of 2D QED (Schwinger model)  for 
this explanation.
We consider the case in which $\Del\Ga$ consists of local counterterms.
\footnote{
In Ref.\cite{CD01}, the divergence structure in 2D QED is closely examined.
For the bosonic part of the effective action, we need not consider
the non-perturbative divergences. Furthermore, by using the
special regularization (Jackiw-Rajaramann parameter $a=1$), 
we can calculate using the fermion measure, 
where the Wess-Zumino term does not appear.
}
From the {\it power-counting} analysis,
\footnote{
The mass dimensions of the gauge (photon) field $A_\m$ and the gauge coupling
(electric charge) $e$
are $0$ and $1$, respectively.
}
 we have
\begin{eqnarray}
\Del\Ga[A]=\int d^2x \left(
c_{\mbox{div}} e^2 {A_\m}^2
+c'_{\mbox{div}}  e \,\pl_\m A_\m
+c''_{\mbox{div}} e \,\ep_\mn \pl_\m A_\n
\right)\com
\label{REN2}
\end{eqnarray}
where $c_{\mbox{div}},c'_{\mbox{div}}$ and 
$c''_{\mbox{div}},$ are some (divergent) constants to be
systematically fixed. Now, we apply the following 
{\it renormalization condition} 
to $\Ga_{\mbox{bare}}[A]$:
\begin{eqnarray}
\left.\frac{\del^2\Ga_{\mbox{bare}}[A]}{\del A_\m(x)\del A_\n(y)}\right|_{A=0}
=\frac{e^2}{\pi}\del_\mn \del^2(x-y)\ ,\ 
\left.\frac{\del \Ga_{\mbox{bare}}[A]}{\del A_\m(x)}\right|_{A=0}
=0\pr
\label{REN3}
\end{eqnarray}
The first equation defines the renormalized coupling, and
the second one guarantees stability.
\footnote{
It is known that 2D (massless) QED is exactly solvable
in the non-perturbative treatment. (See Ref.\cite{AAR91} for
a good textbook.) Here, we treat it perturbatively
in order to compare with 4D QED in \S\S 4 and 5.
The first condition of (19) implies that 2D QED is a massive
vector theory with mass $\frac{\mbox{e}}{\sqrt{\pi}}$.
Here we include information from the exact result.
}

In order to regularize the $t$-integral in $\Ga[A]$, 
we introduce here two more {\it regularization parameters}, 
$\ep$ and $T$\ ($0<\ep<t<T\ ,\ \ep\ra +0\ ,\ T\ra +\infty$)
{\bf [Regularization Prescription 4]}:
\begin{eqnarray}
\Ga[A]=-\ln\,Z[A]=\nn
\lim_{T\ra +\infty,\ep\ra +0}\lim_{M\ra 0}
\int_\ep^{T}\frac{dt}{t}
\half(1-i\frac{\pl}{\pl(tM)})\Tr ( \GfMp(x,y;t)+
                                   \GfMm(x,y;t) )
  \pr
\label{REN4}
\end{eqnarray}
$T$ is regarded as the length of the extra axis, and 
$\ep$ is the ``regularized point'' of the origin of the extra axis.
It can also be regarded as the minimum unit of length
(the ultraviolet cutoff). 
$T$ and $\ep$ regulate 
the infrared and  ultraviolet behaviors, respectively.
The relevant part is the order $A^2$ term,
\begin{eqnarray}
\GfMp|_{AA}
=\int^{X^0}_0dZ^0\int^{Z^0}_0dW^0\int d^2Z\int d^2W
(G^p_0(X,Z)-G^n_0(X,Z))ie\Aslash(z)  \nn
(G^p_0(Z,W)-G^n_0(Z,W))ie\Aslash(w)
(G^p_0(W,Y)-G^n_0(W,Y))             \nn
=\int^{X^0}_0dZ^0\int^{Z^0}_0dW^0\int d^2z\int d^2w   
\int\frac{d^2k}{(2\pi)^2}\frac{d^2l}{(2\pi)^2}\frac{d^2q}{(2\pi)^2}\nn
\times (-i)(\Om_+(k) e^{-iE(k)(X^0-Z^0)}
           -\Om_-(k) e^{iE(k)(X^0-Z^0)})
ie\Aslash(z)                             \nn
\times (-i)(\Om_+(l) e^{-iE(l)(Z^0-W^0)}
           -\Om_-(L) e^{iE(l)(Z^0-W^0)})
ie\Aslash(w)                             \nn
\times (-i)(\Om_+(q) e^{-iE(q)(W^0-Y^0)}
           -\Om_-(q) e^{iE(q)(W^0-Y^0)})
 e^{-ik(x-z)-il(z-w)-iq(w-y)}
\ .
\label{DW13c}
\end{eqnarray}
(See Fig.1(ii).) Evaluating the above equation 
and the similar one for $\GfMm|_{AA}$, we obtain
\begin{eqnarray}
\Ga[A]|_{AA}=\lim_{T\ra +\infty,\ep\ra +0}\frac{e^2}{\pi}
\ln\frac{T}{\ep}\times 
\int d^2x {A_\m}^2\pr
\label{REN6}
\end{eqnarray}
From the renormalization condition (\ref{REN3}), we obtain
\begin{eqnarray}
c_{\mbox{div}}=-\frac{1}{\pi}\ln\frac{T}{\ep}+\frac{1}{2\pi}
\pr
\label{REN7}
\end{eqnarray}

We now check above result by calculating the Weyl and chiral anomalies
{\it directly from the effective action}. 
(In (\ref{DWR12}) and (\ref{DWR13}) of this paper, and in Ref.\cite{SI99}, 
anomalies are obtained from
{\it Jacobians}.)
The Weyl anomaly is obtained
from the scale transformation of $\ep$ (or $T$),
\begin{eqnarray}
\ep'=e^{\Del}\ep=\ep+\Del\cdot\ep+O(\Del^2)\com\nn
\del\Ga_{\mbox{bare}}=\del\, [c_{\mbox{div}}e^2 \int d^2x {A_\m}^2]
=\Del\times\frac{e^2}{\pi}\int d^2x {A_\m}^2
\com
\label{REN8}
\end{eqnarray}
which agrees with the known result. 
The chiral anomaly is obtained by the variation of $A_\m$, 
$\del\Aslash=-\frac{1}{e}\dslash\al\cdot\gago$, for 
$\Ga_{\mbox{bare}}[A]$
\footnote{
The 2 dimensional QED, $\Lcal=\psibar i(\dslash+i e\Aslash)\psi$, 
is invariant under the {\it local} chiral gauge transformation\ 
$\psi'=e^{i\al(x)\gago}\psi\ ,\ \Aslash'=\Aslash
-\frac{1}{e}\dslash\al\cdot\gago.
$
(
$A'_\m=A_\m+\frac{i}{e}\ep_\mn \pl_\n\al,\ \ep_{12}=1.$
)

}
:
\begin{eqnarray}
\frac{\del \Ga_{\mbox{bare}}[A]}{\del \al(x)}=+\frac{i}{\pi} e \ep_\mn\pl_\m A_\n
\pr
\label{REN9}
\end{eqnarray}
Thus, both Weyl and chiral anomalies are correctly reproduced. 
\footnote{
We notice a contrasting point in the two anomalies.
The Weyl anomaly does not depend on the renormalization condition (\ref{REN3}),
whereas the chiral one does depend on it. The former gives a response
to the scale change, and hence picks up the divergent part proportional
to $\ln T/\ep$, while the latter gives a response
to the (chiral) phase change, and hence the charge normalization,
which is defined by the renormalization condition (\ref{REN3}),
is crucial for it. 
}
Note that the chiral
anomaly derived from the Jacobian in Ref. \cite{SI98} 
comes from the $O(A)$-part of $G^{5M}_\pm$,
whereas that derived here from the effective action comes from the 
$O(A^2)$-part. 
(As for
the Weyl anomaly, the results obtained from 
both approaches come from $O(A^2)$-part.)

We conclude this section by listing all regularization
parameters introduced and comparing them with the situation in lattice.
We have introduced {\it three} parameters $M, T$ and $\ep$. They should
satisfy the following relations with two {\it configuration variables},\ 
the 4-momentum $k^\m$ and the inverse-temperature $t$\ :
\begin{eqnarray}
|k^\m|\leq M\ll \frac{1}{t}\com\q\frac{1}{T}<\frac{1}{t}<\frac{1}{\ep}\com\q
\frac{1}{T}\ll \frac{1}{\ep}
\pr
\label{REN10}
\end{eqnarray}
In the above,
we clearly see an important feature of the parameter $M$.
Among the three relations above, the latter two are rather familiar.
(The UV cut-off scale is $\ep$, while the IR cut-off scale is $T$.)
The interesting one is the first.
Before the appearance of domain wall regularization, 
we do not know of such a regularization parameter that depends on the
configuration ( specified by $k^\m$ and $t$ in the present case)
in this way.
In the conclusion of this paper, we point out another important
character of the present regularization, which is obtained from
the result above and that of the next section.
In the lattice domain wall case, {\it four}
parameters are introduced:\ $m_0,a,L_s,l$.\cite{Vra98} 
The correspondence with the present case is as follows:
\begin{eqnarray}
\mbox{present paper} &                     &\mbox{Ref.\cite{Vra98}}\nn
M & \leftrightarrow & m_0\ (\mbox{1+4 dim fermion mass}),\nn
\ep  & \leftrightarrow & a\ (\mbox{lattice spacing}), \nn
\frac{T}{\ep} & \leftrightarrow & L_s\ 
(\mbox{the total site number along the extra axis}),\nn
T  & \leftrightarrow &l\ (\mbox{physical extent of one direction
of the 4 dimensional box}).\nn
\label{REN11}
\end{eqnarray}
The one extra parameter in the lattice  comes from the fact that
the extra space is treated independently of the 4 dimensional space. 
It is adopted in the ordinary domain wall formulation, whereas
the present one treats the extra space as the space of 
the {\it inverse temperature}, 
which appears in expressing the determinant. 
In Ref.\cite{Vra98}, the 4 dimensional fermion mass $m_f$ is also introduced. 
In such a case, we also introduce the mass parameter.
(See Ref.\cite{SI99})

\section{General treatment of renormalization}
To this point, we have discussed only the fermion determinant
for the external gauge field.
In order for this approach to be regarded as 
an alternative new regularization for general field theories, 
the quantum treatment of the gauge fields should also be described.
We devote this section to this task.

The present approach quite naturally fits in 
the {\it background field method}.\cite{tH73,IO82}
We explain it by again taking 4D Euclidean QED as an example. 
The quantum effects of 
both the gauge (photon) field and the fermion are taken into account.  
We consider the massive fermion,
\begin{eqnarray}
\Lcal[\psi,\psibar,A]=i \psibar\{ \ga_\m(\pl_\m+i e A_\m)+im\}\psi
-\fourth {F_\mn}^2-\half (\pl_\m A_\m)^2
\com
\label{GF1}
\end{eqnarray}
where the Feynman gauge is taken.
According to the general theory of the background field method,
complete physical information is contained in the following
background effective action:
\begin{eqnarray}
e^{-\Ga[\chi,\chibar,A]}=\int\Dcal a_\m \Dcal\psi\Dcal\psibar
\exp \left[-\int d^4x\{ \Lcal[\chi+\psi,\chibar+\psibar,A+a]\right.\nn
\left.
-\frac{\del\Lcal[\Phi]}{\del\Phi^i}\phi^i-\Lcal[\chi,\chibar,A]\}\right]\nn
=\int\Dcal a_\m \Dcal\psi\Dcal\psibar\exp 
\left[
-\int d^4x\{\,
\Lcal_2[\psi,\psibar,a]+\Lcal_3[\psi,\psibar,a]\,\}
\right]   \nn
\Lcal_2[\psi,\psibar,a;\chi,\chibar,A]=
i \psibar\{ \ga_\m(\pl_\m+i e A_\m)+im\}\psi 
-\half (\pl_\m a_\n)^2- e (\psibar\aslash\chi+\chibar\aslash\psi)\com\nn
\Lcal_3[\psi,\psibar,a]=-e\psibar \aslash\psi 
\pr\nn
\label{GF2}
\end{eqnarray}
Here, 
the fields $(\Phi_i)=(\chi,\chibar,A_\m)$ are the {\it background} fields, 
and $(\phi_i)=(\psi,\psibar,a_\m)$ are the {\it quantum} fields.
We note that the term $\Lcal_3=-e\psibar\aslash\psi$ is the third order in 
the quantum fields and contributes to orders of
2-loops and higher.
Terms in $\Lcal_2$ are all second order and contribute to 
only the 1-loop order.
Among them, the two terms 
$- e \psibar\aslash\chi$ and $- e \chibar\aslash\psi$ 
are off-diagonal with respect to the quantum fields. 

In order to diagonalize the 1-loop part, 
we redefine the quantum fields as 
$\psi'=\psi+\Del\psi,\psibar'=\psibar+\Del\psibar,
a_\m'=a_\m+\Del a_\m$.
Here we choose $\Del\psi,\Del\psibar$ and $\Del a_\m$
to be {\it linear} with respect to the quantum fields in order to 
{\it maintain the loop-order structure}.
We require the 1-loop part, $\Lcal_2[\psi,\psibar,a]$, 
to be equal to 
\begin{eqnarray}
i \psibar'\{ \ga_\m(\pl_\m+i e A_\m)+im\}\psi'
-\half (\pl_\m {a_\n}')^2
\pr
\label{GF3}
\end{eqnarray}
Then $\Del\psi$ and $\Del\psibar$ should satisfy
\begin{eqnarray}
\Del\psibar (-i\leftdslash-e\Aslash-m)=-e\chibar\aslash\com\nn
(i\dslash-e\Aslash-m)\Del\psi=-e\aslash \chi
\pr
\label{GF4}
\end{eqnarray}
From this, we know that $\Del\psi$ and $\Del\psibar$ are 
proportional to the vector quantum field $a_\m$ and begin from 
order $e$. Then $\Del a_\m$ satisfies
\begin{eqnarray}
\int d^4x\{
\half (\pl_\m a_\n+\pl_\m (\Del a_\n))^2- \half (\pl_\m a_\n)^2\}
=
\int d^4x\{
-\half e \,\Delta\psibar\aslash\chi-\half e\,\chibar\aslash\Delta\psi\}
\ ,\nn
\label{GF4b}
\end{eqnarray}
where the relations (\ref{GF4}) have been used.
Then, we see that the solution $\Del a_\m$ is obtained
as the expansion in the coupling $e$, beginning from 
order $e^2$. [The RHS of (\ref{GF4b}) begins from order of $e^2$.]
We have 
\begin{eqnarray}
\Del a_\m=e^2X^{(2)}_\m+e^3X^{(3)}_\m+O(e^4)
\pr
\label{GF5}
\end{eqnarray}
Assuming $a_\n\pl_\m X^{(2)}_\n$
damps sufficiently rapidly at the boundary $|x_\m|=\infty$,
$X^{(2)}_\m$ satisfies the equation
\begin{eqnarray}
\pl^2 X^{(2)}_\m=\frac{1}{2e}(\chibar\ga_\m\Del\psi|_e+
\Del\psibar|_e\ga_\m\chi)
\pr
\label{GF6}
\end{eqnarray}
Here,``$|_e$'' represents the operation of extracting the 
first order part in $e$. Therefore the RHS is 0-th order
in $e$.

Equation (\ref{GF4}) can be perturbatively  solved by requiring the natural
condition that $\Del\psi$ and $\Del\psibar$ vanish when
the quantum fields vanish, $(a_\m,\psi,\psibar)=0$. This gives
\begin{eqnarray}
\Del\psi(x)=-\int d^4y\,S^A(x-y) e\aslash(y)\chi(y)
=-\int d^4y\,S(x-y) e\aslash(y)\chi(y)+O(e^2)\ ,\nn
\Del\psibar(x)=-\int d^4y\, e\chibar(y)\aslash(y)\Sbar^A(x-y)
=-\int d^4y\, e\chibar(y)\aslash(y)\Sbar(x-y)+O(e^2)\ ,
\label{GF7}
\end{eqnarray}
where $S^A, \Sbar^A, S$ and $\Sbar$ are defined by
\begin{eqnarray}
(i\dslash- e\Aslash-m)S^A(x-y)=\del^4(x-y)\ ,\ 
(i\dslash-m)S(x-y)=\del^4(x-y)\ ,\nn
\Sbar^A(x-y)(-i\leftdslash- e\Aslash-m)_y=\del^4(x-y)\ ,\ 
\Sbar(x-y)(-i\leftdslash-m)_y=\del^4(x-y)\ .
\label{GF8}
\end{eqnarray}
Here $S^A(x-y)$ and $\Sbar^A(x-y)$ are background dependent
propagators. 
All these propagators are (4D) Euclidean ones, and there
is no ambiguity in the choice of boundary conditions.

The path-integral expression (\ref{GF2}) can be rewritten by
using redefined quantum fields using the measure change
\begin{eqnarray}
\Dcal a_\m \Dcal\psi \Dcal\psibar=J\times
\Dcal {a_\n}'\Dcal\psi' \Dcal\psibar'\com\q
J\equiv\frac{\pl (a_\m,\psi,\psibar)}{\pl ({a_\n}',\psi',\psibar')}
\pr
\label{GF9}
\end{eqnarray}
Since we keep the linear relation in the choice of
the redefined quantum fields,
the Jacobian $J$ does {\it not} depend on quantum fields:
\begin{eqnarray}
e^{-\Ga[\chi,\chibar,A]}
=J\times \int\Dcal {a_\m}' \Dcal\psi'\Dcal\psibar'\exp [-\int d^4x\{
i \psibar'\ga_\m(\pl_\m+i e A_\m+im)\psi'  \nn
-\half (\pl_\m {a_\n}')^2+\mbox{cubic terms w.r.t. quantum fields}\}]
\com
\label{GF9B}
\end{eqnarray}
The 1-loop part in the integrand is the quantity
 discussed in \S 2 for the case $m=0$ ( \S 7 of Ref.\cite{SI99}
treats the case $m\neq 0$).
\footnote{
The cubic terms in (\ref{GF9B}) can be explicitly written
in terms of $(\psi',\psibar',{a_\m}')$ using (\ref{GF5}), (\ref{GF6})
and (\ref{GF7}). They involve non-local interactions.
We can calculate higher-loop parts using the cubic terms. 
}
Up to the lowest nontrivial order, $J$ is obtained as
\begin{eqnarray}
J^{-1}=\frac{\pl ({a_\m}',\psi',\psibar')}{\pl (a_\n,\psi,\psibar)}
=\det (\del_\mn\del^4(x-y)+e^2\frac{\del X^{(2)}_\m(x)}{\del a_\n(y)}+
O(e^3))  \nn
=\exp\, [e^2 \Tr\,(\frac{\del X^{(2)}_\m(x)}{\del a_\n(y)})+O(e^3)]\ ,\nn
\ln J=e^2\Tr_{x=z,\m=\n}\int d^4y D(x-y)
\chibar(y)\ga_\m S(y-z)\ga_\n\chi(z)+O(e^3)
\pr
\label{GF10}
\end{eqnarray}
This corresponds to the  1-loop part for the fermion self-energy. 
(See Fig. 2.)
\begin{figure}
\centerline{\epsfysize=3cm\epsfbox{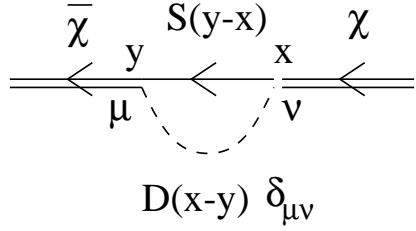}}
\caption{
Fermion self-energy.}
\end{figure}
Indeed, the quantum effect of the gauge field is taken into
account. 
(Demonstrating this was the original motivation of the present work.)
Because the Jacobian is {\it decoupled} from the main integral part,
the regularization for the divergences in (\ref{GF10})
can be done {\it independently} of the regularization parameter $M$
introduced in Sec.2. 
The quantity to be regularized in Sec.2 was
$\det (\dslash+i e\Aslash)$,
while the Jacobian above is roughly 
$\det (1+e^2\frac{1}{\pl^2}\chibar\frac{1}{\dslash}\chi)=
\det (\pl^2+e^2\chibar\frac{1}{\dslash}\chi)/\det\pl^2$.
Both quantities are divergent, but
the presence of ``$\pl^2$'' inside the determinant 
in the latter quantity causes
no chiral problem. 
[There is no ambiguity in determining the divergent quantity
$\det (\pl^2+\cdots)$ because it can be expressed as the
heat equation (not the Dirac equation ) in 
1+4 dimension. (See \S 2 of Ref.\cite{SI99}.)
]
The (momentum) integral, corresponding to  ``trace''  in (\ref{GF10}),  
is evaluated in the usual way, 
as described in standard field theory textbooks.

We explain the effective action  calculation in the {\it coordinate} space
in more detail. We have
\begin{eqnarray}
\ln J|_{e^2}=e^2\int d^4y \int d^4x D(x-y)
\chibar(y)\ga_\m S(y-x)\ga_\m\chi(x)\nn
=e^2\int d^4y \int d^4x D(x-y)      
\chibar(y)\ga_\m S(y-x)\ga_\m\{
\chi(y)+(x-y)^\m\pl_\m\chi|_y+\cdots
\}\nn
\equiv\Ga_{\chibar\chi}+\Ga_{\chibar\pl\chi}+\cdots
\com\nn
\label{GF11}
\end{eqnarray}
where $\chi(x)$ is Taylor expanded around $y$, and each expanded
part is defined by $\Ga_{\chibar\chi}$,$\Ga_{\chibar\pl\chi}$,$\cdots$.
Here $D(x)$ comes from the photon propagation and is defined by
$\pl^2D(x)=\del^4(x)$. 
The usual integral calculation gives the (ultra) divergent parts as 
\begin{eqnarray}
\Ga_{\chibar\chi}=\frac{e^2}{2\pi^2}m\ln\frac{T}{\ep}\times
\int d^4y\,\chibar(y)\chi(y)+\mbox{finite term}\com\nn
\Ga_{\chibar\pl\chi}=-
\frac{e^2}{8\pi^2}\ln\frac{T}{\ep}\times
\int d^4y\,i\chibar(y)\dslash \chi(y)+\mbox{finite term}
\com
\label{GF12}
\end{eqnarray}
where we take the region of the momentum integral as 
$\frac{1}{T}<|k_\m|<\frac{1}{\ep}$, where $T$ and $\ep$ were
introduced in \S 3 as the infrared and ultraviolet cutoffs.
Here, we find the fermion part of the counter action as 
[the gauge part is given in (\ref{REN2}) for 2D QED] 
\begin{eqnarray}
\Del\Ga[\chi,\chibar]=\int d^4y\,\left\{
\frac{e^2}{2\pi^2}m\ln\frac{T}{\ep}\times
\chibar(y)\chi(y)-\frac{e^2}{8\pi^2}\ln\frac{T}{\ep}\times
i\chibar(y)\dslash \chi(y)       \right\}
\ ,
\label{GF13}
\end{eqnarray}
which is introduced in order to cancel the divergences of (\ref{GF12}).
The quantity $\Del\Ga$ is ``absorbed'' by the mass renormalization 
$m+\Del m$ and the wave function
renormalization of the fermion $\sqrt{Z_2}$ as follows: 
\begin{eqnarray}
\Del\Ga[\chi,\chibar,A]\equiv \int d^4x\Del\Lcal[\chi,\chibar,A]\com\nn
\Lcal[\chi,\chibar,A;e,m]+\Del\Lcal[\chi,\chibar,A]
=\Lcal[\sqrt{Z_2}\chi,\sqrt{Z_2}\chibar,\sqrt{Z_3}A;e+\Del e,m+\Del m]\com\nn
m+\Del m=m(1-\frac{3e^2}{8\pi^2}\ln\frac{T}{\ep})\com\q
\sqrt{Z_2}=1-\frac{e^2}{16\pi^2}\ln\frac{T}{\ep}
\pr\nn
\label{GF14}
\end{eqnarray}
Note in particular that
 the mass is renormalized in the {\it multiplicative} manner, 
as expected.
\footnote{
In the present continuum approach, this multiplicative
renormalization results from the fact 
that the Jacobian $J$ decouples from the $M$-involved part
(fermion determinant part) in (\ref{GF9B}). In the lattice, the 4D fermion
mass is essentially introduced as the coupling $m_1$ between the walls
at the two ends of the extra axis.
In this case, determining 
whether or not the renormalized mass is proportional to $m_1$ 
is highly non-trivial, because $M$ could appear  additively, simply
for the dimensional reasons. A numerical simulation supports the
validity of the
multiplicative renormalization. This fact is very important
for there to be no need for 
fine-tuning (good control of the small mass fermion)
and for the validity of the chiral perturbation.\cite{Sha93}
}


\section{Discussion and conclusion}
Combining the results of \S\S 3 and 4, we present the final form
of the renormalization prescription (taking 4D QED as an example)
with the background field gauge:\cite{IO82}
\begin{eqnarray}
(\Phi_i)=(\chi,\chibar,A):\ \mbox{Background fields}\com\q
(\p_i)=(\psi,\psibar,a):\ \mbox{Quantum fields}\com\nn
\Lcal[\chi,\chibar,A;e,m]=i \chibar\ga_\m(\pl_\m+i e A_\m+im)\chi
-\fourth {F_\mn}^2\com\nn
\Lcal_{\mbox{gauge}}[a;\xi]=-\frac{1}{2\xi} (\pl_\m a_\m)^2\com\nn
e^{-\Ga_{\mbox{bare}}[\Phi]}=\int\Dcal \p\,
\exp \left[-\int d^4x\{ \Lcal[\Phi+\p]+\Lcal_{\mbox{gauge}}[a]
+\Del\Lcal[\Phi+\p]\right.\nn
\left. -\frac{\del}{\del\Phi^i}(\Lcal[\Phi]+\Del\Lcal[\Phi])\phi^i
-\Lcal[\Phi]-\Del\Lcal[\Phi]\}\right]
\pr\nn
\label{CON1}
\end{eqnarray}
Here, the fermion determinant part only is regularized by
the domain wall regularization of \S 2, while other parts
are regularized 
by the cutoffs for the $t$-axis:
$\ep<t<T (\mbox{or }\frac{1}{T}<|k_\m|<\frac{1}{\ep})$.
The background field gauge is chosen here and $\xi$ is the
gauge parameter. 
$\Del\Lcal$ is obtained in such a way that $\Ga_{\mbox{bare}}$
becomes finite, satisfying 
some proper
{\it renormalization condition} on the following quantities:
\begin{eqnarray}
\left.\frac{\del^2\Ga_{\mbox{bare}}[\chi,\chibar,A]}
{\del A_\m(x)\del A_\n(y)}\right|_{\chi=\chibar=A=0}
\ ,\ 
\left.\frac{\del^2\Ga_{\mbox{bare}}[\chi,\chibar,A]}
{\del \chi(x)\del \chibar(y)}\right|_{\chi=\chibar=A=0}
\ ,\ \nn
\left.\frac{\del^3\Ga_{\mbox{bare}}[\chi,\chibar,A]}
{\del A_\m(x)\del \chi(y)\del \chibar(z)}\right|_{\chi=\chibar=A=0}
\pr
\label{CON2}
\end{eqnarray}
In paticular, the fermion mass and the gauge coupling
are normalized by the second and the third conditions, 
respectively.
The renormalization parameters are obtained by
\begin{eqnarray}
\Lcal[\chi,\chibar,A;e,m]+\Del\Lcal[\chi,\chibar,A]
=\Lcal[\sqrt{Z_2}\chi,\sqrt{Z_2}\chibar,\sqrt{Z_3}A;
e+\Del e,m+\Del m]
\pr
\label{CON3}
\end{eqnarray}
Comparing with the case in \S 4, we have the relation
\begin{eqnarray}
\sqrt{Z_3}=\frac{1}{1+\frac{\Del e}{e}}
\com
\label{CON4}
\end{eqnarray}
because the background gauge invariance is preserved. 
Some comments, in relation to the higher-loop structure, are in order.
\cite{IO82}
\begin{enumerate}
\item
Generally, the terms in the Taylor expansion of $\Del\Lcal$ play the role
of subtracting sub-divergences in multi-loop diagrams.
The proof of this point 
is largely based on the structure of the Taylor expansion.
\item
From the viewpoint of the Taylor expansion, we compare the treatment
of the gauge part in this section and \S 4.
In \S 4, the gauge term $-\half (\pl_\m A_\m)^2$ is properly
Taylor expanded in (\ref{GF2}). Therefore, in this case, the subdivergence
problem is manifestly solved. Contrastingly, 
the background field gauge adopted in this section,
$\Lcal_{\mbox{gauge}}$ in (\ref{CON1}), is not Taylor expanded.
The proof of the subdivergence cancellation is carried out
by using the properties described in item 3, below. 
The superiority of (\ref{CON1}) is that the effective action
$\Ga_{\mbox{bare}}$ is guaranteed to be gauge invariant.
\item
In relation to the subdivergence problem mentioned above,
at orders of 2-loops and higher, 
the gauge parameter and quantum fields suffer from
the renormalization effect 
(i.e., the renormalization of ``internal'' quantities). 
\end{enumerate}

Through this analysis, the character of the domain wall
regularization is revealed. 
The first point of note 
is the condition on the regularization parameters (\ref{REN10}), 
as stated in \S 3. The second point is that
the treatment of the fermion loop (determinant) 
is different from that of the other types of loops, 
which are irrelevant to the chiral problem. 
For the fermion loop, we do the calculation in the following order:\ 
1) the momentum ($k^\m$) integral in the region $|k^\m|\leq M$
is evaluated for fixed $t$;\  
2) the procedure $Mt\ll 1$ is carried out;\  
3) the extra coordinate ($t$) integral is evaluated. 
Therefore, for each $t$-segment, the momentum region is suppressed
as $|k_\m|\leq M_t (\ll \frac{1}{t})$, where the subscript $t$
indicates a possible slight $t$ dependence.
\footnote{
The consistency with the 1+4dim Dirac equation (\ref{DWR4})
requires $|t\pl_tM|\ll M$. This implies that 
$M=\mbox{const}\times t^{\pm \th},\ $ with $0<\th\ll 1$
} 
In order to show the restricted configuration region,
 we present a schematic
chart of the momentum-integration region in 
a ``phase space" ($|k^\m|,t$). (See Fig.3.)
\begin{figure}
\centerline{\epsfysize=7cm\epsfbox{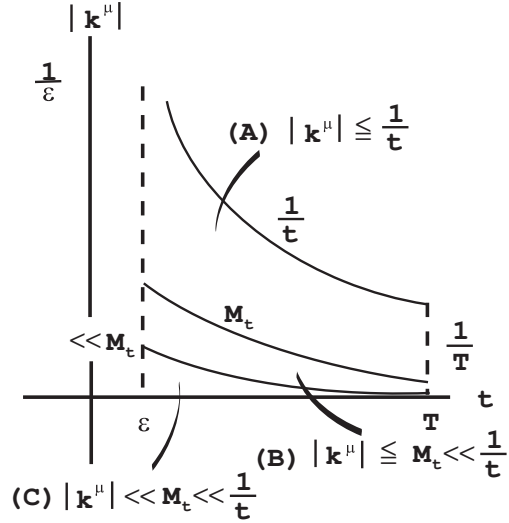}}
\caption{
The restricted regions of the momentum-integral
in the ``phase space" ($|k^\m|,t$).
All relations between parameters and the momentum
are listed in (\ref{REN10}). 
Three regions are shown:\ 
(A)\ [below the $t^{-1}$ line] the ordinary integral region;\ 
(B)\ [below the $M_t$ line] the domain wall regularization;\ 
(C)\ [far below the $M_t$ line] the extreme chiral region.}
\end{figure}
The present regularization considers 
the region below the $M_t$ line. 
For other types of loops, the corresponding
region is suppressed as $|k_\m|\leq \frac{1}{t}$. 
(In Fig.3, this is the region below the $t^{-1}$ line.)
Two upper cut-offs
have the relation $M_t\ll\frac{1}{t}$ from the requirement (\ref{REN10}). 
It is expected that 
the final physical results are not affected by 
the different treatments, 
depending on the types of loops, 
of the momentum integrals 
as far as the {\it low energy} fermion
and gauge bosons are concerned.
(This situation is realised in the lattice numerical simulation.)

In the literature to this time, the domain wall has been investigated 
mainly with regard to the determinant calculation
for the external gauge field.
In the present paper, we have formulated it 
in the general field theory framework,
 using the background field method, where both
fermions and (gauge) bosons are treated as quantum fields. 
We can calculate 
any term, in principle, of the effective action. 
We have explicitly carried out the
renormalization  of the fermion
wave function and the fermion mass. We find that 
they agree with previous results.

Finally, we comment on the relation with the recently
popular higher-dimensional approaches, such as
the brane-world approach, Randall-Sundrum model, etc.
They start from a higher-dimensional gravitational theory
and consider a soliton configuration localized
in the extra space. In 
this setting, the dimensional
reduction from 5 dimensions to 4 dimensions takes place. 
In accordance with this, the mass hierarchy
characterized by some exponential factor
appears. This makes the model building based
on this mechanism attractive. 
Contrastingly, the present approach
starts from the 4 dimensions and for the purpose of 
the chiral regularization, we make use of 
its 5 dimensional Dirac equation.
We suspect that both approaches do essentially the same thing, 
based on the following similar ingredients involved: 
1) the domain wall configuration;\ 
2) the exponentially damping or growing factors;\ 
3) the chiral properties;\ 
4) the scaling role played by the extra-space parameter or coordinate;\ 
5) the fact that the regularization parameter $M$ in the present treatment
appears to correspond to the parameter of (thickness)$^{-1}$
in the RS-model;\ 
6) the fact that infrared-ultraviolet relation appears.
The purpose of the higher-dimensional models is to find
a theory beyond the standard model. This is different from the present 
purpose, namely, 
regularization of the fermion system. We can, however, find
a common root in the paper of Callan and Harvey\cite{CH85}.

\section*{Acknowledgements}
This work began at the Albert Einstein Institut
(Max Planck Institut, Potsdam)
in the autumn of 1999. 
The author thanks H. Nicolai for reading the initial manuscript 
and for his hospitality there.
He also thanks the governor of Shizuoka prefecture for financial support.
This work was completed in the present form at the Research Institute
for Mathematical Sciences (Kyoto University, Kyoto) in the autumn of 2001.
The author thanks I. Ojima for reading the manuscript and
the members of the institute for their hospitality.

%

\end{document}